\documentclass[twocolumn,showpacs,preprintnumbers,amsmath,amssymb]{revtex4}

\usepackage{graphicx,epsfig}
\usepackage{dcolumn}
\usepackage{bm}
\def\cA{{\cal A}}

\begin{document}

\preprint{}

\title{Equilibration through local information exchange in networks}

\author{K. Y. Michael Wong$^1$}
\author{David Saad$^2$}

\affiliation{ $^1$Department of Physics, The Hong Kong University
of Science and Technology, Clear Water Bay, Hong Kong, China
\\ $^2$The Neural Computing Research Group, Aston University,
Birmingham B4 7ET, United Kingdom}

\date{\today}

\begin{abstract}
We study the equilibrium states of energy functions involving a
large set of real variables, defined on the links of sparsely
connected networks, and interacting at the network nodes, using
the cavity and replica methods. When applied to the representative
problem of network resource allocation, an efficient distributed
algorithm is devised, with simulations showing full agreement with
theory. Scaling properties with the network connectivity and the
resource availability are found.
\end{abstract}


\maketitle

Many theoretically challenging and practically important problems
involve interacting variables connected by network
structures~\cite{Nishimori_book}. Statistical mechanics of
disordered systems makes contributions towards the understanding
of such systems at two levels. Macroscopically, it describes the
typical behavior of the networks, using techniques such as the
replica method. Microscopically, it analyzes the relation between
variables, using techniques such as the cavity method, that give
rise to efficient computational algorithms. In computer science,
probabilistic inference based on graphical structures has been
developed and
applied~\cite{pearl1988,mackaybook}, mainly for providing
approximate solutions to specific instances of limited size.
Examples of recent success included the belief propagation
algorithm for error-correcting codes~\cite{os} and the survey
propagation algorithm for the satisfiability
problem~\cite{mezard}.

Most analyses so far have focused on networks of discrete
variables. However, many typical problems, such as network
resource allocation, involve continuous variables. Compared with
discrete variables, analyses for continuous variables were much
less explored. The main obstacle comes from the need to pass among
the nodes entire free energy functions as messages. This is much
more complex than cases of discrete values, where the messages are
countable sets of conditional probability estimates of {\it
discrete values}. Previous work in the computer science literature
focused on modeling these functions for getting good
approximations in feasible time scales~\cite{lauritzen1996}. There
have been attempts to simplify the messages for continuous
variables, for example, to parametrize them using eigenfunction
decomposition for special cases, but the general feasibility
remains an open question \cite{skantzos}.

In this Rapid Communication we study a system of real variables
that can be mapped onto a sparse graph. Based on the analysis, we
demonstrate the close relationship between belief propagation
algorithms and the Bethe approximation in statistical
physics~\cite{YWF}, and propose a message-passing
approximation method, generally applicable to problems of
continuous variables. The method is efficient since the messages
consist of only the first and second derivatives of the vertex
free energies derived from our analysis. The key to the successful
simplification, not needed for the simpler case of discrete
variables, is that the messages passed to a target node
are accompanied by information-provision messages
from the target node, to first determine the state at which the
derivatives should be calculated.

We first formulate the problem at a general temperature, and then
focus on a prototype for optimization. The traditional approach
for optimization on networks is to adopt computationally demanding
global optimization techniques, such as linear or quadratic
programming~\cite{bertsekas}. In contrast, message-passing
approaches have the potential to solve global optimization
problems via local updates, thereby reducing the computational
complexity. An even more important advantage, relevant to
practical implementation, is its distributive nature. Since it
does not require a global optimizer, it is particularly suitable
for distributive control in large or evolving networks.

We consider a sparse network with $N$ nodes, labeled $i=1,\dots,N$.
Each node $i$ is randomly connected to $c$
other nodes. The connectivity matrix is given by $\cA_{ij}=1, 0$
for connected and unconnected node pairs respectively. A link
variable $y_{ij}$ is defined on each connected link from $j$ to
$i$. We consider an energy function (cost)
$E\!=\!\sum_{(ij)}\cA_{ij}\phi(y_{ij})
\!+\!\sum_i\psi(\lambda_i,\{y_{ij}|\cA_{ij}\!=\!1\})$,
where $\lambda_i$ is a quenched variable defined on node $i$. In the
context of probabilistic inference, $y_{ij}$ may represent the
coupling between observables in nodes $j$ and $i$, $\phi(y_{ij})$
may correspond to the logarithm of the prior distribution of
$y_{ij}$, and $\psi(\lambda_i,\{y_{ij}|\cA_{ij}\!=\!1\})$ the logarithm
of the likelihood of the observables $\lambda_i$. In the context of
resource allocation, $y_{ij}\!\equiv\!-y_{ji}$ may represent the
current from node $j$ to $i$, $\phi(y_{ij})$ the transportation
cost, and $\psi(\lambda_i,\{y_{ij}|\cA_{ij}\!=\!1\})$ the performance
cost of the allocation task on node $i$, dependent on the node
capacity $\lambda_i$.

We are interested in the case of intensive connectivity $c\!\sim\!
O(1)\!\ll \!N$. Since the probability of finding a loop of finite
length on the network is low, the cavity method well describes the
local environment of a node. A node is connected to $c$ branches
in a tree structure, and the correlations among the branches of
the tree are neglected. In each branch, nodes are arranged in
generations. A node is connected to an ancestor node of the
previous generation, and other $c\!-\!1$ descendent nodes of the
next generation. Considering node $i$ as the ancestor of node $j$,
the descendents of node $j$ form a tree structure ${\mathbf T}$
with $c\!-\!1$ branches, labeled by $k\!\ne\! i$ for
$\cA_{jk}\!=\!1$. At a temperature $T\!\equiv\!\beta^{-1}$, the
free energy $F(y_{ij}|{\mathbf T})$ can be expressed in terms of
the free energies $F(y_{jk}|{\mathbf T}_k)$ of its descendents.
The free energy can be considered as the sum of two parts,
$F(y|{\mathbf T})\!=\!N_{\mathbf T}F_{\rm av}\!+\!F_V(y|{\mathbf
T})$, where $N_{\mathbf T}$ is the number of nodes in the tree
${\mathbf T}$, $F_{\rm av}$ is the average free energy per node,
and $F_V(y|{\mathbf T})$ is referred to as the {\it vertex free
energy}. This leads to the recursion relation
\begin{eqnarray}
   && \!\!\!\!  F_V(y_{ij}|{\mathbf T})=
    -T\ln\Biggl\{\prod_{k\ne i}\left(\int dy_{jk}\right)
        \exp\biggl[-\beta\psi(\lambda_j,\{y_{jk}\})\nonumber\\
        &&-\beta\sum_{k\ne i}\left(F_V(y_{jk}|{\mathbf T}_k)
    +\phi(y_{jk})\right)\biggr]\Biggr\}\Biggr|_{\cA_{jk}=1}-F_{\rm av},
\label{recurt}
\end{eqnarray}
\begin{eqnarray}
    F_{\rm av}=\!\!\!\!\!
    &&-T\Biggl\langle\ln\Biggl\{
        \prod_k\left(\int dy_{jk}\right)
        \exp\biggl[-\beta\psi(\lambda_j,\{y_{jk}\})\nonumber\\
        &&-\beta\sum_k\left(F_V(y_{jk}|{\mathbf T}_k)
    +\phi(y_{jk})\right)\biggr]\Biggr\}
    \Biggr|_{\cA_{jk}=1}\Biggr\rangle_\lambda,
\label{free}
\end{eqnarray}
where ${\mathbf T}_k$ is the tree terminated at node $k$, and
$\langle\dots\rangle_\lambda$ represents the average over the
distribution of $\lambda$.
Interestingly, the recursive relation of Eq.~(\ref{recurt}) can be
directly linked to probabilistic message passing (belief propagation),
where the logarithms of messages passed between nodes are proportional to
the vertex free energies.

For more concrete discussions, we focus
on a prototype for optimization,
termed resource allocation
and well known in the areas of computer science
and operations management~\cite{resourceallocation,resourceallocation2}.
The analysis of the problem is applicable to typical situations where
a large number of nodes are required to balance loads and/or resources, such
as reducing internet traffic congestion and streamlining network flows
of commodities~\cite{Shenker}. In computer science, many practical
solutions are usually heuristic
and focus on practical aspects (e.g., communication protocols).
Here we study a more generic version of the problem
represented by nodes of some
computational power that should carry out tasks. Both
computational powers and tasks will be chosen at random from some
arbitrary distribution. The nodes are located on a randomly chosen
sparse network of some connectivity. The goal is to allocate tasks
on the network such that demands will be satisfied while the
migration of (sub-)tasks is minimized.

We focus here on the satisfiable case where the total computing
power is greater than the demand, and where the number of nodes
involved is very large; the unsatisfiable case can be investigated
using a similar approach~\cite{WS_long}. Each node on the network
has a capacity (computational capability minus allocated tasks)
$\lambda_i$ randomly drawn from a distribution $\rho(\lambda_i)$.
With the aim to satisfy the capacity constraints, we have
$\psi(\lambda_i,\{y_{ij}|\cA_{ij}\!=\!1\}) \!=\!\ln[\Theta(-\sum_j
\cA_{ij}y_{ij}\!-\!\lambda_i)\!+\!\epsilon]$, where $\epsilon\to
0$. The problem reduces to the load balancing task of minimizing
the energy function (cost)
$E\!=\!\sum_{(ij)}\cA_{ij}\phi(y_{ij})$, subject to the capacity
constraints $\sum_j\cA_{ij}y_{ij}\!+\!\lambda_i\ge 0$.

When $\phi(y)$ is a general even function of the current $y$,
we may also derive Eq.~(\ref{recurt}) using the replica method.
We first introduce the {\em chemical potentials} $\mu_i$ of nodes $i$,
and approximate the current $y_{ij}$
as driven by the potential differences between nodes
$y_{ij}\!=\!\mu_j\!-\!\mu_i$.
Since sparse networks are locally treelike,
the probability of finding short loops is vanishing in large networks,
and the approximation works well.

Considering the optimization problem in the space of chemical potentials,
we calculate the replicated partition function
$\langle Z^n\rangle_{\cA,\lambda}$
averaged over the connectivity matrix and capacity distribution,
and take the limit $n\!\to\! 0$.
Assuming replica symmetry, the saddle point equations
yield a recursion relation for a two-component function $R$
dependent on the tree structure ${\mathbf T}$, given by
\begin{eqnarray}
    &&\!\!R(z,\mu|{\mathbf T})=
    \frac{1}{\cal D}\prod_{k\!=\!1}^{c-1}\left(
        \int d\mu_k R(\mu,\mu_k|{\mathbf T}_k)\right)
        \Theta\!\Biggl(\sum_{k\!=\!1}^{c-1}\mu_k\!\nonumber\\
    &&\!-\!c\mu\!+\!z\!+\!\lambda_{V({\mathbf T})}
        \!\Biggr)\exp\!\left(\!-\frac{\beta\epsilon}{2}\mu^2
        -\beta\sum_{k\!=\!1}^{c-1}\phi(\mu-\mu_k)\!\right)\!,
\label{recurr}
\end{eqnarray}
where $\cal D$ is a constant, ${\mathbf T}_k$ represents the tree
terminated at the $k$th descendent, and
$\lambda_{V({\mathbf T})}$ the capacity of the vertex of the tree
${\mathbf T}$. The term $\beta\epsilon\mu^2/2$, with
$\epsilon\!\to\!0$, is introduced to break the translational
symmetry of the chemical potentials, since the energy function is
invariant under the addition of a constant to all chemical
potentials.

Equation (\ref{recurr}) expresses $R(z,\mu|{\mathbf T})$ in terms of
$c\!-\!1$ functions $R(\mu,\mu_k|{\mathbf T}_k)$
($k\!=\!1,..,c\!-\!1$), a characteristic of the tree structure.
Furthermore, except for the factor $\exp(-\beta\epsilon\mu^2/2)$,
$R$ is a function of $y\!\equiv\!\mu\!-\!z$,
which is interpreted as the current
drawn from a node with chemical potential $\mu$ by its ancestor
with chemical potential $z$. One can then express the function $R$
as the product of a {\em vertex partition function} $Z_V$ and a
normalization factor $W$, that is, $R(z,\mu|{\mathbf
T})\!=\!W(z)Z_V(y|{\mathbf T})$. In the limit $\epsilon\!\rightarrow\!0$,
the dependence on $\mu$ and $y$ decouples,
enabling one to derive a recursion relation
for the {\em vertex free energy}
$F_V(y|{\mathbf T})\!\equiv\!-\!T\ln Z_V(y|{\mathbf T})$
and arrive at Eq.~(\ref{recurr}).

The current distribution and the average free energy per link can
be derived by
integrating the current $y'$ in a link from one vertex to another,
fed by the trees ${\mathbf T}_1$ and ${\mathbf T}_2$,
respectively; the obtained expressions are
$P(y)\!=\!\langle\delta(y-y')\rangle_{\star}$ and $\langle
E\rangle\!=\!\langle\phi(y')\rangle_{\star}$ where
$
    \langle \bullet\rangle_{\star}
    =\!\left\langle
        \int dy'\exp\left[-\beta E(y')\right](\bullet)/
        \int dy'\exp\left[-\beta E(y')\right]
    \right\rangle_\lambda,
$
and $E(y')\!=\!F_V(y'|{\mathbf T}_1)\!+\!F_V(-y'|{\mathbf T}_2)\!+\!\phi(y')$.

Figure~1(a) shows results of the iteration of Eq.~(\ref{recurt}),
in the case of optimization ($T\!=\!0$) based on discretizing
$F_V(y|{\mathbf T})$ as a vector. The capacity distribution
$\rho(\lambda)$ is Gaussian of variance 1 and average
$\langle\lambda\rangle$. Each iteration corresponds to adding one
extra generation (1000 new nodes in our simulations) to the tree
structure, such that the iterative process corresponds to
approximating the network by an increasingly extensive tree.  We
observe that after an initial rise with iteration steps, the
average energies converge to steady-state values, at a rate which
increases with the average capacity.

To study the convergence rate of the iterations, we fit the
average energy at iteration step $t$ using $\langle
E(t)\!-\!E(\infty)\rangle\!\sim\!\exp(-\gamma t)$ in the
asymptotic regime. As shown in the inset of Fig.~1(a), the
relaxation rate $\gamma$ increases with the average capacity.
A cusp appears at the average capacity of about 0.45, below
which convergence is slow due to a plateau that develops in the
average energy curve before the final stage.
The slowdown is probably due to the appearance of increasingly large
clusters of nodes with negative resources, 
which draw currents from increasingly extensive regions of nodes with
excess resources to satisfy the demand.

\begin{figure}[t]
\begin{center}
\begin{picture}(220,290)
\put(0,150){\epsfxsize=67mm  \epsfbox{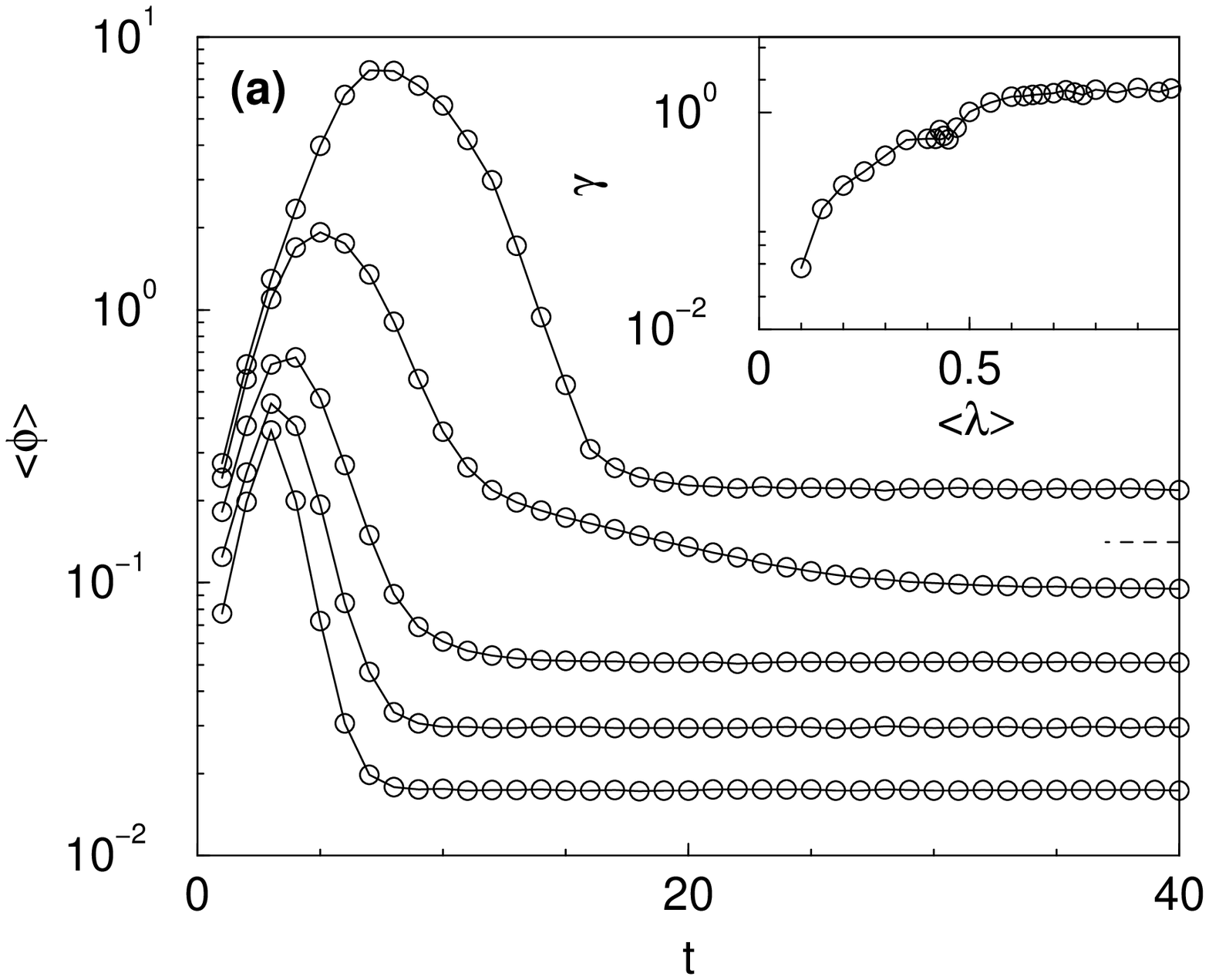}}
\put(0,-10){\epsfxsize=67mm  \epsfbox{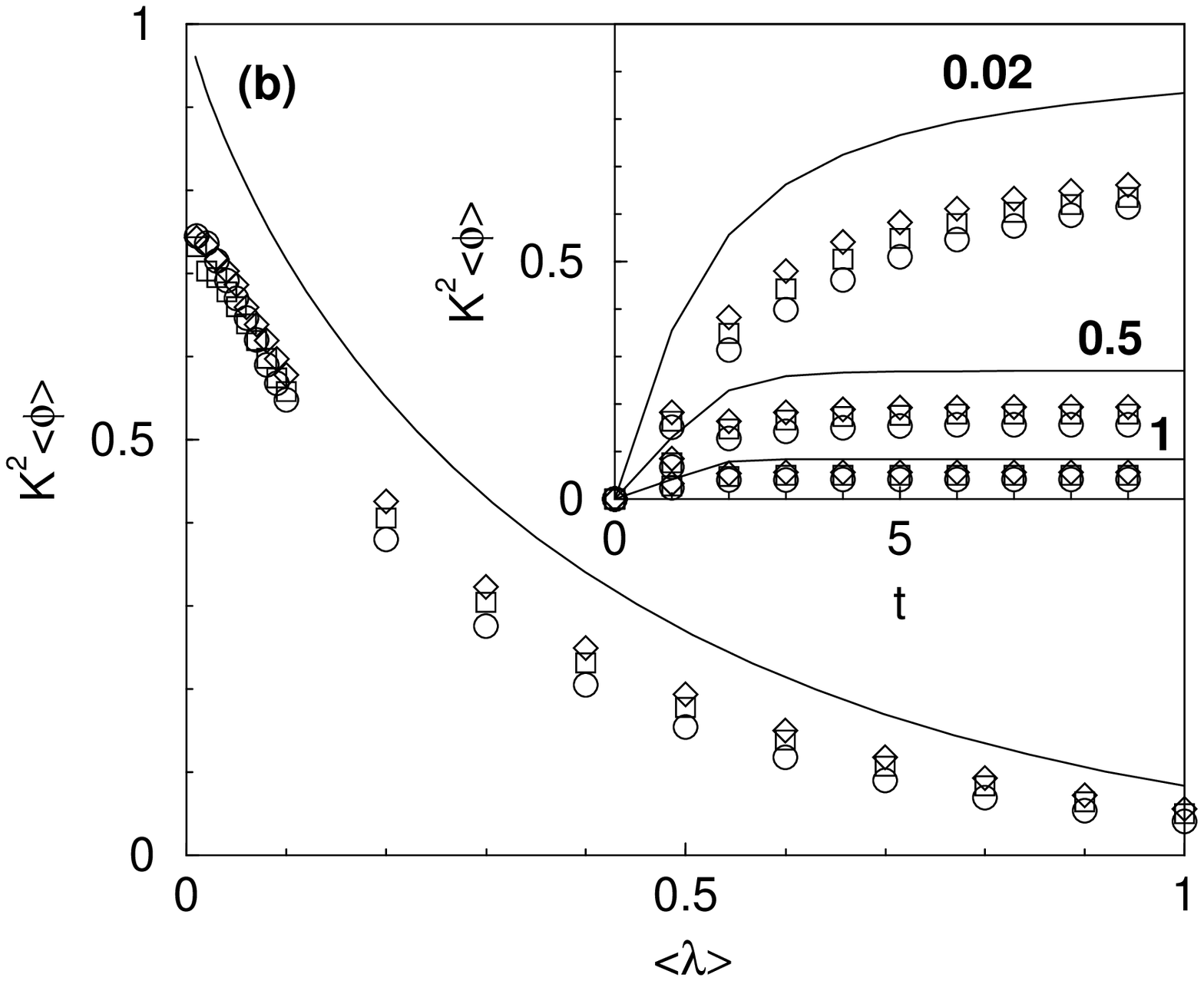}}
\end{picture}
\caption{Results for $N\!=\!1000$ and $\phi(y)=y^2/2$.
(a) $\langle\phi\rangle$ obtained by iterating Eq.~(\ref{recurt}) as
a function of $t$ for $\langle\lambda\rangle\!=\!$ 0.1, 0.2, 0.4,
0.6, 0.8 (top to bottom) and $c\!=\!3$. Dashed line: The
asymptotic $\langle\phi\rangle$ for $\langle\lambda\rangle\!=\!0.1$.
Inset: $\gamma$ as a function of $\langle\lambda\rangle$.
(b) $K^2\langle\phi\rangle$ as a function of $\langle\lambda\rangle$
for $c=3$ ($\bigcirc$), 4 ($\square$), 5 ($\lozenge$), large $K$ (line).
Inset: $K^2\langle\phi\rangle$ as a function of time
for random sequential update of Eqs.~(\ref{msgab}) and (\ref{msgmu}).
Symbols: same as (a).
\vspace*{-1cm}}
\label{figure1}
\end{center}
\end{figure}
\begin{figure}[t]
\begin{center}
\begin{picture}(220,290)
\put(0,150){\epsfxsize=67mm  \epsfbox{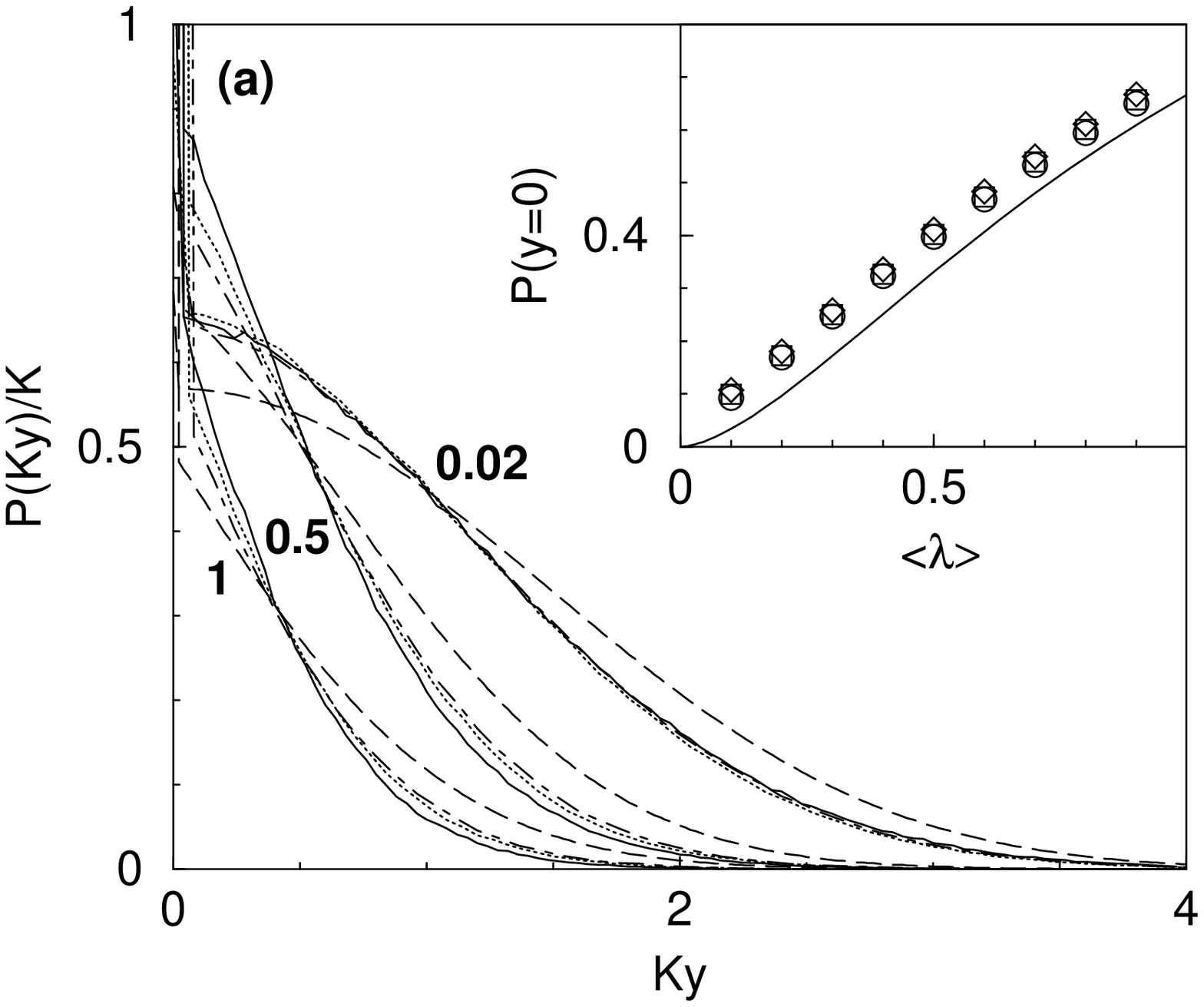}}
\put(0,-10){\epsfxsize=67mm  \epsfbox{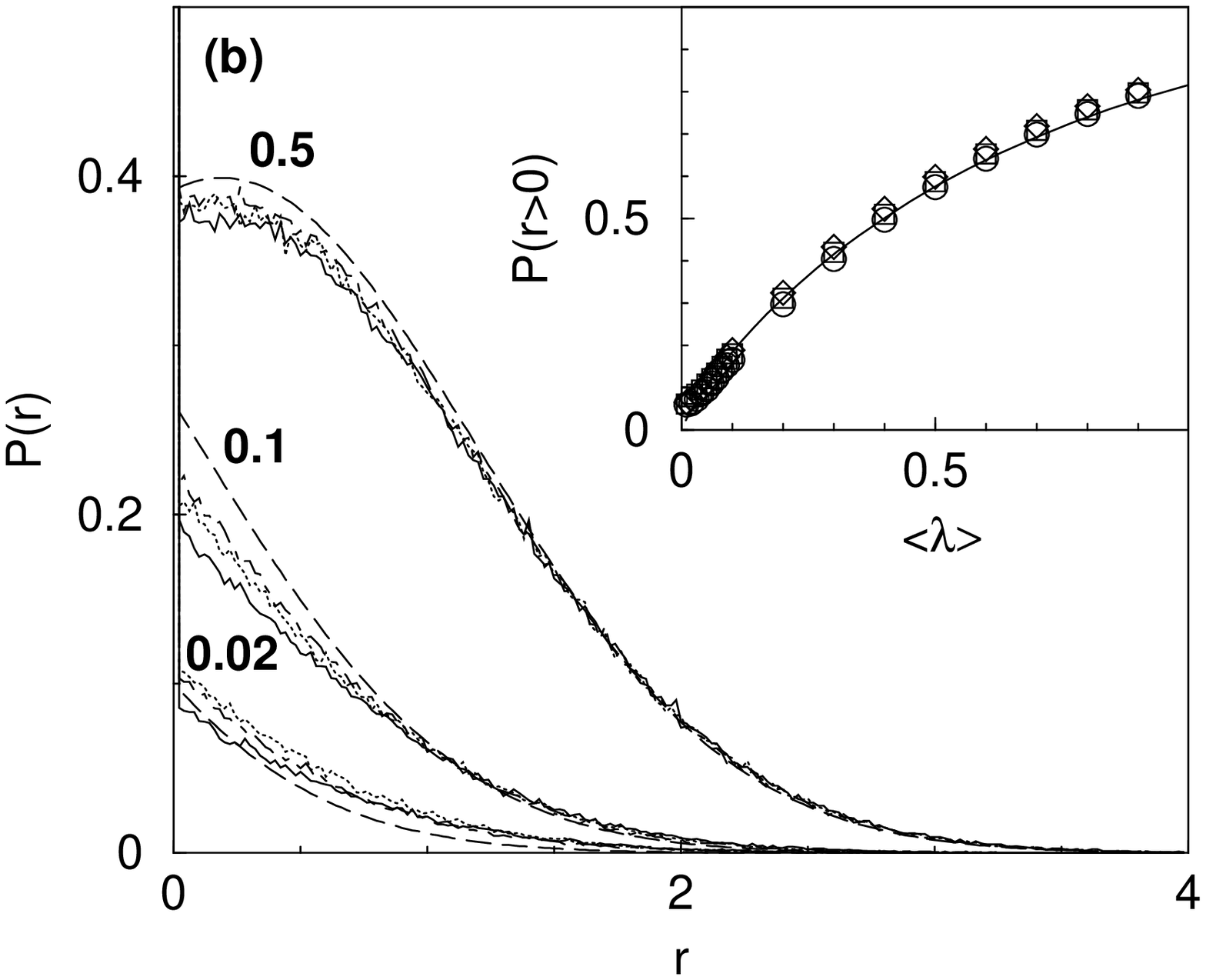}}
\end{picture}
\caption{Results for $N\!=\!1000$ and $\phi(y)\!=\!y^2/2$. (a) The
current distribution $P(Ky)/K$ for
$\langle\lambda\rangle\!=\!0.02, 0.5, 1$, and $c\!=\!3$ (solid
lines), 4 (dotted), 5 (dot-dashed), large $K$ (long dashed).
Inset: $P(y\!=\!0)$ as a function of $\langle\lambda\rangle$ for
$c\!=\!3$ ($\bigcirc$), 4 ($\square$), 5 ($\lozenge$), large $K$
(line). (b) The resource distribution $P(r)$ for
$\langle\lambda\rangle\!=\!0.02, 0.1, 0.5$, large $K$. Symbols: as
in (a). Inset: $P(r\!>\!0)$ as a function of
$\langle\lambda\rangle$. Symbols: as in the inset of (a).
\vspace*{-0.5cm}} \label{figure2}
\end{center}
\end{figure}

The local nature of the recursion relation~(\ref{recurt}) points to
the possibility that the network optimization can be solved by
message-passing approaches.
Instead of passing the {\it functions} $F_V(y|{\mathbf T})$
of the current $y$ as messages,
we simplify each message to two parameters,
namely, the first and second derivatives of the vertex free energies.
Let \!$(A_{ij},B_{ij})\!\equiv \![\partial F_V(y_{ij}|{\mathbf
T}_j)/\partial y_{ij},
\partial^2 F_V(y_{ij}|{\mathbf T}_j)/\partial y_{ij}^2]$
be the message passed from node $j$ to $i$. Using
Eq.~(\ref{recurt}), the recursion relations
lead to the message $(A_{ij},B_{ij})$
\begin{equation}
    A_{ij}\!\leftarrow\!-\mu_{ij},
        \quad
        B_{ij}\!\leftarrow\!\frac{\Theta(-\mu_{ij}+\epsilon)}
    {\sum_{k\ne i}\cA_{jk}(\phi''_{jk}+B_{jk})^{-1}},
\label{msgab}
\end{equation}
\begin{eqnarray}
    &&\mu_{ij}\!=\!\min\Biggl\{
    \Biggl[\sum_{k\ne i}\cA_{jk}[y_{jk}-(\phi'_{jk}+A_{jk})
    (\phi''_{jk}+B_{jk})^{-1}]
    \nonumber\\
    &&+\lambda_j-y_{ij}\Biggr]
    \Biggl[\sum_{k\ne i}\cA_{jk}(\phi''_{jk}+B_{jk})^{-1}\Biggr]^{-1},
    0\Biggr\},
\label{msgmu}
\end{eqnarray}
with $\phi'_{jk}$ and $\phi''_{jk}$ representing
the first and second derivatives of $\phi(y)$
at $y=y_{jk}$, respectively.

The algorithm is complete
with the determination of the drawn current $y_{ij}$
at which the derivatives comprising the messages should be computed.
Two methods are proposed.
In the first, when messages are sent from node $j$ to the ancestor
node $i$, backward messages $y_{jk}$ computed from the same
optimization step are sent from node $j$ to the descendent nodes
$k$, informing them of the particular arguments to be used for
calculating subsequent messages. In the second, node $j$ first
receives the messages $(A_{ji},B_{ji})$ and current $y_{ji}$ {\em
from} the ancestor node $i$, and update the current $y_{ij}$ by
minimizing the total cost. Both methods work well for the
quadratic cost functions.

For comparison, an independent exact optimization is available 
at zero temperature. The chemical potentials turn out to be
the Lagrange multipliers of the capacity constraints, and the
relation between the currents and the
chemical potentials turns out to be exact. The K\"uhn-Tucker
conditions for the optimal solution yield
\begin{equation}
        \mu_i\!=\!\min\left[\frac{1}{c}\left(
        \sum_j  \cA_{ij}\mu_j+\lambda_i\right),0\right].
\label{condit}
\end{equation}
Like in the message-passing algorithm, this condition also
provides a local iterative solution to the optimization problem.
Simulations show that it yields excellent agreement
with Eqs.~(\ref{recurt}), (\ref{msgab}), and (\ref{msgmu}).

To study the dependence on the connectivity,
we first consider the limit of large $K\!\equiv \!c\!-\!1$,
in which Eq.~(\ref{msgab}) converges to the steady-state results of
$A_{ij}\!=\!\max[K^{-1}\sum_{k\ne i}\cA_{jk}A_{jk}\!-\!\lambda_j),0]$
and $B_{ij}\!\sim\! K^{-1}$.
Then $\sum_{k\ne i}\cA_{jk}A_{jk}$ becomes self-averaging
and equal to $Km_A$,
where $m_A\!\sim \!K^{-1}$ is the mean of the messages $A_{ij}$.
Thus, $y_{ij}\!\sim\!\mu_i\!\sim\! K^{-1}$.
The physical picture of this scaling behavior
is that the current drawn by a node
is shared among the $K$ descendent nodes.
After rescaling, quantities
such as $K^2\langle\phi\rangle$, $P(Ky)/K$ and $P(K\mu)/K$
become purely dependent on the average capacity $\langle\lambda\rangle$.

For increasing finite values of $K$,
Fig. 1(b) shows the common trend of $K^2\langle\phi\rangle$
decreasing with $\langle\lambda\rangle$ exponentially,
and gradually approaching the large $K$ limit.
The scaling property extends to the optimization dynamics
[Fig. 1(b) inset].
As shown in Fig.~2(a),
the current distribution $P(Ky)/K$ consists of a $\delta$ function
component at $y\!=\!0$ and a continuous component,
whose breadth decreases with $\langle\lambda\rangle$.
Remarkably, the distributions for different connectivities
collapse almost perfectly after the currents are rescaled by $K^{-1}$,
with a very mild dependence on $K$
and gradually approaching the large $K$ limit.
As shown in the inset of Fig.~2(a),
the fraction of idle links increases with $\langle\lambda\rangle$.
The fraction has a weak dependence on the connectivity,
confirming the almost universal distributions rescaled for different $K$.

Since the current on a link scales as $K^{-1}$,
the allocated resource of a node
should have a weak dependence on the connectivity.
Defining the resource at node $i$ by
$r_i\!\equiv \lambda_i\!+\!\sum_j\cA_{ij}y_{ij}$,
the resource distribution $P(r)$ shown in Fig.~2(b)
confirms this behavior even at low connectivities.
The fraction of nodes with unsaturated capacity constraints
increases with the average capacity,
and is weakly dependent on the connectivity [Fig.~2(b) inset].
Hence the saturated nodes form a percolating cluster
at a low average capacity,
and breaks into isolated clusters at a high average capacity.
It is interesting to note that at
the average capacity of 0.45, below which a plateau starts to develop
in the relaxation rate of the recursion relation, Eq.~(\ref{recurt}),
the fraction of saturated nodes is about 0.47, close to the
percolation threshold of 0.5 for $c\!=\!3$.

In summary, using the example of the resource allocation problem
on sparsely connected networks,
we studied the use of message-passing methods for equilibration
using both replica and cavity based analyses. A local
algorithm was devised and successfully applied to this task.
The study also reveals the scaling properties
of this model,
showing that the resource distribution on the nodes
depends principally on the networkwide availability of resources,
and depends only weakly on the connectivity.
Links share the task of resource provision,
leading to current distributions that are
almost universally dependent on the resource availability
after rescaling.

While the analysis focused on fixed connectivity and zero
temperature, it can accommodate any connectivity profile and
temperature parameter and may be used for analyzing a range of
inference problems. For instance, we have considered the effects
of adding anharmonic and frictional terms to the quadratic cost
function. The message-passing function can be adapted to these
variations, and the results will be presented
elsewhere~\cite{WS_long}. Both analysis and algorithm extend the
use of current message-passing techniques to inference in problems
with continuous variables, opening up a rich area for further
investigations with many potential applications.

We thank David Sherrington for useful comments.
This work is partially supported by the Research Grant
Council of Hong Kong (Grants HKUST6062/02P, 
DAG04/05.SC25, and DAG05/06.SC36)
and EU Grants EVERGROW, IP 1935 in FP6 and STIPCO in FP5.


\begin{thebibliography}{0}

\bibitem{Nishimori_book}
H. Nishimori, {\it Statistical Physics of Spin Glasses and Information
Processing} (Oxford University Press, Oxford, UK, 2001).

\bibitem{pearl1988}
J. Pearl, {\it Probabilistic Reasoning in Intelligent Systems}
(Morgan Kaufmann, San Mateo, CA, 1988).

\bibitem{mackaybook}
D. J. C. Mackay, {\it Information Theory, Inference and Learning
Algorithms}, (Cambridge University Press, Cambridge, UK, 2003).

\bibitem{os}
M. Opper and D. Saad, {\it Advanced Mean Field Methods}
(MIT Press, Cambridge, MA, 2001).


\bibitem{mezard}
M. M\'ezard, e-print cond-mat/0401237.

\bibitem{lauritzen1996}
S. L. Lauritzen, {\it Graphical Models}
(Oxford University Press, New York, 1996).

\bibitem{skantzos}
N. Skantzos, I. P. Castillo, and J. P. L. Hatchett, 
Phys. Rev. E {\bf 72}, 066127 (2005).

\bibitem{YWF} J. S.~Yedidia, W. T.~Freeman, and Y.~Weiss, IEEE Trans. 
Inf. Theory \textbf{51}, 2282 (2005).

\bibitem{bertsekas}
D. Bertsekas, {\it Linear Network Optimization} 
(MIT Press, Cambridge, MA, 1991).

\bibitem{resourceallocation}
L. Peterson and B. S. Davie, {\it Computer Networks: A Systems Approach}
(Academic Press, San Diego, CA, 2000).

\bibitem{resourceallocation2}
Y. C. Ho, L. Servi, and R. Suri, Large Scale Syst. {\bf 1}, 51 (1980).

\bibitem{Shenker}
S. Shenker, D. Clark, D. Estrin, and S. Herzog, ACM
Comp.~Comm.~Rev. {\bf 26} 19 (1996).



\bibitem{WS_long}K . Y. M. Wong and D.~Saad unpublished.

\end{thebibliography}
\end{document}